\documentclass{article}

\usepackage{graphicx}
\usepackage{fullpage,amsmath,amssymb,latexsym}
\usepackage{multirow}
\usepackage{natbib}

\begin{document}

\title{What do we Really Know about Uranus and Neptune?}
\author{M. Podolak and R. Helled\\
\small{Dept. of Geophysical, Atmospheric, and Planetary Sciences}\\ 
\small{Raymond and Beverly Sackler Faculty of Exact Sciences}\\
\small{Tel Aviv University, Tel Aviv, Israel}\\
}

\date{}
\maketitle

\begin{abstract}
The internal structures and compositions of Uranus and Neptune are not well constrained due to the uncertainty in rotation period and flattening, as well as the relatively large error bars on the gravitational coefficients. While Uranus and Neptune are similar in mass and radius, they differ in other physical properties such as thermal emission, obliquity, and inferred atmospheric enrichment. In this letter we consider the uncertainty in the planetary rotation periods, show that rotation periods more consistent with the measured oblateness imply that Uranus and Neptune have different internal structures, and speculate on the source of that difference.  We conclude that Uranus and Neptune might have very different structures and/or compositions despite their similar masses and radii.  We point out that understanding these differences can have important implications for our view of the formation and evolution of Uranus and Neptune as well as intermediate-mass extra-solar planets in general. 
\end{abstract}

\section{Introduction}
Uranus and Neptune are often thought of as twin planets. They both formed in the outer solar system, and their masses and radii are very similar, as are their gravitational moments.  Even their rotation periods are within 10\% of each other. Yet, there are important differences between these planets. Uranus' mass is slightly smaller than Neptune's, but its radius is somewhat larger, so the difference in mean density, is considerable. Indeed interior models of Uranus and Neptune (Podolak et al. 1995; Helled et al. 2011) indicate that Neptune's outermost envelope is more enriched in high-Z material than that of Uranus. 
 
In addition, Neptune has an internal heat source, while Uranus is in equilibrium with solar insolation (Pearl et al. 1990; Pearl \& Conrath 1991) suggesting that Uranus' interior may not be fully convective, and/or that it contains compositional gradients which hinder convection. The difference in thermal flux is indicated from calculations of Uranus' thermal evolution which find that unlike Neptune, Uranus cannot reach its measured intrinsic luminosity within solar system's age if an adiabatic interior is assumed (Fortney \& Nettelmann 2010, Fortney et al., 2011). Another distinct feature of Uranus is its large axial tilt. This was almost certainly caused by a dramatic event in the early history of the solar system (Safronov 1966), although an alternative explanation for Uranus' tilt is orbital migration ({Bou\'{e}} \& {Laskar}, 2010).  Stevenson (1986) suggested that this same dramatic event might have caused the necessary compositional gradients for inhibiting convection. We return to this point later.
  
Despite the available measurements, the internal structures of Uranus and Neptune are not well constrained, and as a result, their bulk compositions are essentially unknown (Podolak et al. 1995; Marley et al. 1995; Helled et al. 2011). In fact, it is still unclear what the mass fraction of water is in Uranus and Neptune, despite their categorization as `icy planets'. \par

Recently, Nettelmann et al. (2012), showed that by assuming modified solid-body rotation periods for Uranus and Neptune, the derived internal structures can change considerably. It was concluded  that an uncertainty in the planetary rotation period can be crucial for constraining the internal structure. In this letter, we investigate further how the uncertainties in the gravity field, shape, and rotation period might affect the inferred planetary interior. 

\section{Observed Parameters}
Interior models use observed physical properties to constrain the planetary composition and its depth dependence. Below we summarize the measurements and suggest that their uncertainties, when included in interior models, can be crucial for understanding the formation, evolution and interiors of the planets. 

\subsection{Gravitational Moments}
The gravitational moments are probably the best known physical properties of Uranus and Neptune after their masses. Although the error bars are fairly large and only $J_2$, $J_4$ are known, the gravity data provide tight constraints on the planetary density profile. Recently, Jacobson (Jacobson 2007, 2009) re-determined the gravitational harmonics Uranus and Neptune and provided more accurate estimates for the gravitational harmonics with significantly smaller error bars. The updated gravitational data of Uranus and Neptune are given in Table 1.

\subsection{Oblateness}
The planetary oblateness (flattening) is defined by
$$f=\frac{Re-Rp}{Re}$$ where $Re$ and $Rp$ are the equatorial and polar radii, respectively. While knowledge of the planetary continuous shape is available, typically, interior models use only the flattening.  Although stellar and ring occultations provide good determinations of the planetary shape (French et al. 1987), these data correspond to low pressure-levels and might not apply to the 1-bar level that is used by interior modelers since atmosphere dynamics (winds) can change the isobaric shape between pressure levels. The connection between the planetary shape at 1-bar and at lower pressure levels isn't entirely obvious and should be investigated in future research. In addition, there are no direct measurements of the polar radii for Uranus and Neptune, and they are inferred from the measured flattening at lower latitudes.

Lindal et al. (1987) and Lindal (1992) provided estimated the polar and equatorial radii of Uranus and Neptune using occultation data and these values were adopted as the ``standard" for interior models (e.g., Guillot 2005). These shapes, however, were {\it not} inferred from the Voyager 2 radio periods but under the assumption that the planets rotate differentially on cylinders. This explains the inconsistency between the typically used flattening of Uranus and Neptune and the Voyager 2 rotation periods (Helled et al. 2010, hereafter HAS10).  If differential rotation is confined to a thin atmospheric layer, and the deeper interior rotates uniformly, the shape of the geopotential surface is determined by this latter rotation period.  For interior models the shape of the geoid should be used, since it represents most of the mass, but this shape is essentially unknown.

\subsection{Rotation Period}
The rotation periods of Uranus and Neptune are determined from radio and magnetic field data. HAS10 suggested that, based on our experience with Saturn (Gurnett et al. 2007, 2010), the measured radio periods of Uranus and Neptune, 17.24 and 16.11 hr, respectively, might not represent the rotation periods of the deep interior. In addition to radio data, Voyager 2 also provided magnetic field data. For Uranus, Ness et al. (1986) obtained a rotation period of $17.29\pm 0.1$ hr; a period which is close to the radio period. For Neptune, the time base for the flyby was insufficient to determine the period accurately but a rotation period of $\sim 16$ hr was found to fit the magnetic field data fairly well (Ness et al., 1989). Recently, Karkoschka (2011) derived a rotation period of 15.9663$\pm$0.0002 for Neptune by following stationary atmospheric features. This period is only slightly shorter than the Voyager radio period. However, the rotation period of the atmosphere might not be identical to the rotation of the deep interior and therefore should be taken with caution. Finally, the complex nature of the magnetic fields of Uranus and Neptune, and the fact that the location at which the magnetic field is generated, which could be at $\sim 0.7 R_{planet}$ (Stanley \& Bloxham 2006), suggest that the magnetic period might not represent the rotation period of the bulk interior. \par

Although the formal error bars on the measurements of angular speed, $\omega$, and $f$ are fairly small, observational ambiguities make these numbers less certain than the error bars indicate. These two parameters are used together with $J_{2n}$, for inferring the internal structures of Uranus and Neptune, and through them the structures of ice giants among the newly discovered exoplanets.  It is therefore desirable to investigate the implications of possible changes in these parameters since this may have important consequences for our understanding, not only of Uranus and Neptune, but also of a whole class of exoplanets.

\section{The Radau Approximation}
Radau (1885) introduced a useful approximation:  if $\epsilon\equiv\sqrt{f(2-f)}$ is the ellipticity of an equipotential surface in the rotating body, and $a$ is the surface's mean radius, the {\it Radau parameter} is
$$\eta\equiv\frac{a}{\epsilon}\frac{d\epsilon}{da}.$$
Using this parameter and Clairaut's equation, one can show that
$$\frac{d}{da}\left[\bar{\rho}a^5\sqrt{1+\eta}\right]=5\bar{\rho}a^4\psi(\eta)$$
where, if $\rho(a)$ is the density on an equipotential surface, $$\bar{\rho}\equiv\frac{3}{a^3}\int_0^a\rho r^2dr$$ is the average density of the enclosed volume, and 
$$\psi (\eta )=\frac{1+\frac 12\eta -\frac 1{10}\eta ^2}{\sqrt{1+\eta }}.$$
The Radau approximation assumes $\psi(\eta)\equiv 1$.

\subsection{Relations Between Observed Parameters}
The Radau approximation relates the moment of inertia to other, directly measurable, parameters. If $M$ is the planet's mass, $a$ is its mean radius, and $G$ is Newton's constant, we can define a rotation parameter $q\equiv \omega^2R^3/GM$ which is the ratio of the centrifugal to gravitational force.  Using the Radau approximation the moment of inertia, $C$ is related to $q$ and $f$ via (Zharkov \& Trubitsyn 1978).
\begin{equation}\label{c}
\frac C{Ma^2}=\frac 23\left[ 1-\frac 25\left( \frac{5q}{2f}-1\right)
^{1/2}\right]  
\end{equation}
     
For a more precise version of the approximation (Yoder 1995) we set $g(r)\equiv GM(r)/r^2$ and define the constants $\delta_1$ and $\delta_2$ by
\begin{equation}\label{del1}
\delta_1=\frac{\int_0^R x^3g(x)\left[\psi(\eta)-1\right]dx}{\int_0^Rx^3g(x)dx},
\end{equation}
\begin{equation}\label{del2}
\delta_2\approx\frac{3f}{7}+\left[\frac{5q(1-f)}{2f}-2\right]\left[\frac{8q(1-f)-3f}{42}\right].
\end{equation}
The relation between the inertia factor and the observable parameters is given by
\begin{equation}\label{rdyod}
\frac C{Ma^2}=\frac 23\left[ 1-\frac 25\left(\frac{1}{1+\delta_1}\right)\sqrt{\frac{5q(1-f)}{2f}}-1+\delta_2\right].
\end{equation}  
Note that when the Radau approximation is exact ($\psi(\eta)\equiv 1$) $\delta_1=0$.  This isn't precisely true for the solar-system giant planets, but for simplicity we assume $\delta_1=0$. 

The combination of gravitational and rotational forces causes an equipotential surface to be non-spherical.  The requirement that the potential on a {\it level surface} be independent of latitude relates $J_2$, $f$, and $q$.  Yoder (1995) gives
\begin{equation}\label{j2}
J_2\approx\frac 13\left[2f\left(1-\frac{2}{7}f\right)-q(1-f)\left(1-\frac{2}{7}f\right)+\frac{11}{49}qf^2\right].
\end{equation}
\begin{equation}\label{j4}
J_4\approx -\frac{15}{7}J_2^2+\frac{5}{21}\left[\frac 45 f\left(1-\frac 12 f\right)-q(1-f)\left(1-\frac 27 f\right)\right]^2.
\end{equation}
\begin{equation}\label{f}
f\approx\frac{q(1-f)+3J_2}{2}\left(1+\frac 32 J_2\right)+\frac 58 J_4.
\end{equation}
Eq.(\ref{rdyod}) uses the flattening, which has traditionally been regarded as the least reliable of the measured parameters. The usual practice among interior modelers is to assume a solid-body rotation period and use it to match the models to $J_2$ and $J_4$.  However Eqs.\,(\ref{j2}-\ref{f}) can be used to replace either $q$ or $f$ by a combination of $J_2$ and $J_4$.

\begin{figure}
\centerline{\includegraphics[width=13cm]{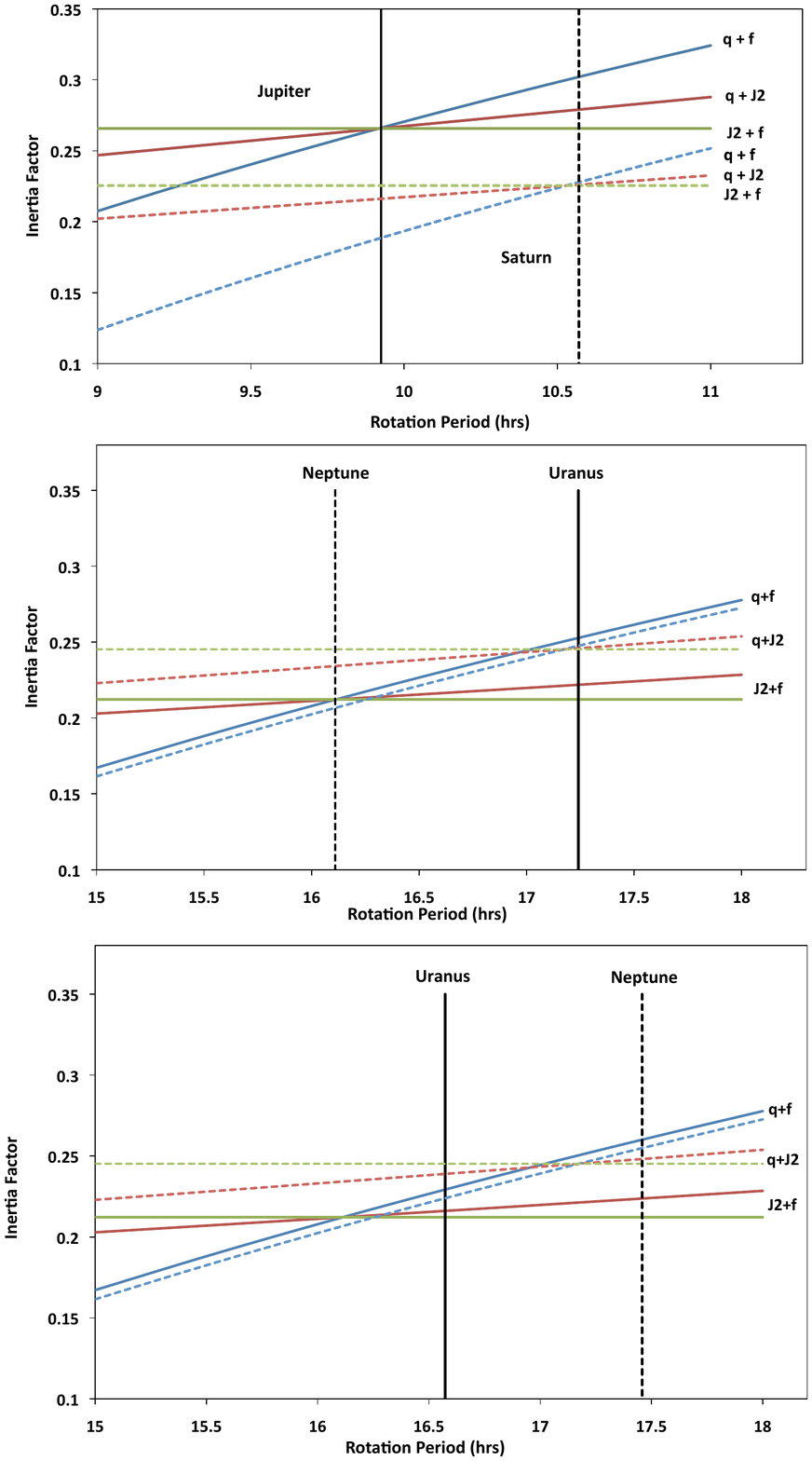}}
\caption{Inertia factor vs. rotation period as computed with three different approximations: $q$ and $J_2$ (maroon), $q$ and $f$ (blue), and $f$ and $J_2$ (green).  The upper panel shows Jupiter (solid) and Saturn (dashed). The measured rotation periods are given by the vertical black lines for Jupiter (solid) and Saturn (dashed).  The middle panel shows the same plots for Uranus (solid) and Neptune (dashed) using the Voyager periods.  The bottom panel is the same, but using HAS10 periods.} 
\label{jands}
\end{figure}

The accuracy of the Radau approximation can be gauged from Fig.\,\ref{jands}.  Plotted is the inertia factor vs. rotation period as determined from one pair of observed parameters: $q$ and $f$ (blue), $q$ and $J_2$ (maroon) and $J_2$ and $f$ (green).  The solid and dotted curves are for Jupiter and Saturn, respectively. The observational parameters used are listed in Table~1, while the inertia factors computed for the different planets and rotation speeds are given in Table~2.  

Since the observational parameters for Jupiter and Saturn are well known, we would expect the three estimates for the inertia factor to coincide at the correct rotation period (shown by the vertical black lines).  As can be seen from the upper panel in the figure, the Radau approximation works extremely well for these two planets.  For Jupiter it's only 35\,s shorter than the system III period.  For Saturn it's 7.72\,m shorter than the Voyager radio period, but only 54\,s shorter than the period determined by Anderson \& Schubert 
(2007). It also suggests that Saturn is more centrally condensed than Jupiter which is consistent with detailed interior models (e.g. Guillot 2005). 
However, when the same quantities are plotted for Uranus and Neptune the disagreement between the period inferred from the Radau approximation and the Voyager radio period is striking as can be seen from the middle panel. Indeed HAS10 pointed out that the Voyager periods may be in error, and suggested rotation periods of 16.57 and 17.46\,hr  for Uranus and Neptune, respectively. 

When we use the Voyager radio periods for $q$, and fit $J_2$ then, as can be seen from the middle panel in Fig.\,\ref{jands}, Uranus and Neptune are found to have inertia factors 0.22 and 0.23, respectively, and their interiors are expected to be rather similar. On the other hand, when we use HAS10 rotation periods the agreement is much better. This is shown in the bottom panel of Fig.\,\ref{jands} which is similar to the bottom panel but with HAS10 rotation periods. The three approximations cross at 16.11\,hr for Uranus and 17.17\,hr for Neptune. HAS10 rotation periods are not precisely equal to those determined from the Radau approximation but clearly in better agreement.  With the modified rotation periods the inertia factor for Uranus is $\sim 0.22$, while the value for Neptune is $\sim 0.25$ implying that Uranus is more centrally condensed than Neptune.  Even the Voyager periods implied a slightly more condensed Uranus, in the sense that it's envelope was closer to solar composition, but the HAS10 periods greatly enhance this difference.

The conclusion that Uranus is more centrally condensed can also be derived from the gravitational moments themselves. The higher order moments are more sensitive to the density distribution in the outer layers of the planet. For planets like Jupiter, where the core is a relatively small fraction of the volume, its contribution to both $J_2$ and $J_4$ is negligible, and the core properties are determined mostly from $J_0$ (i.e., the mass).

\begin{figure}
%\centerline{\includegraphics[width=15cm, angle=-90]{j2_j4_pdf.pdf}}
\centerline{\includegraphics[width=16cm, angle=0]{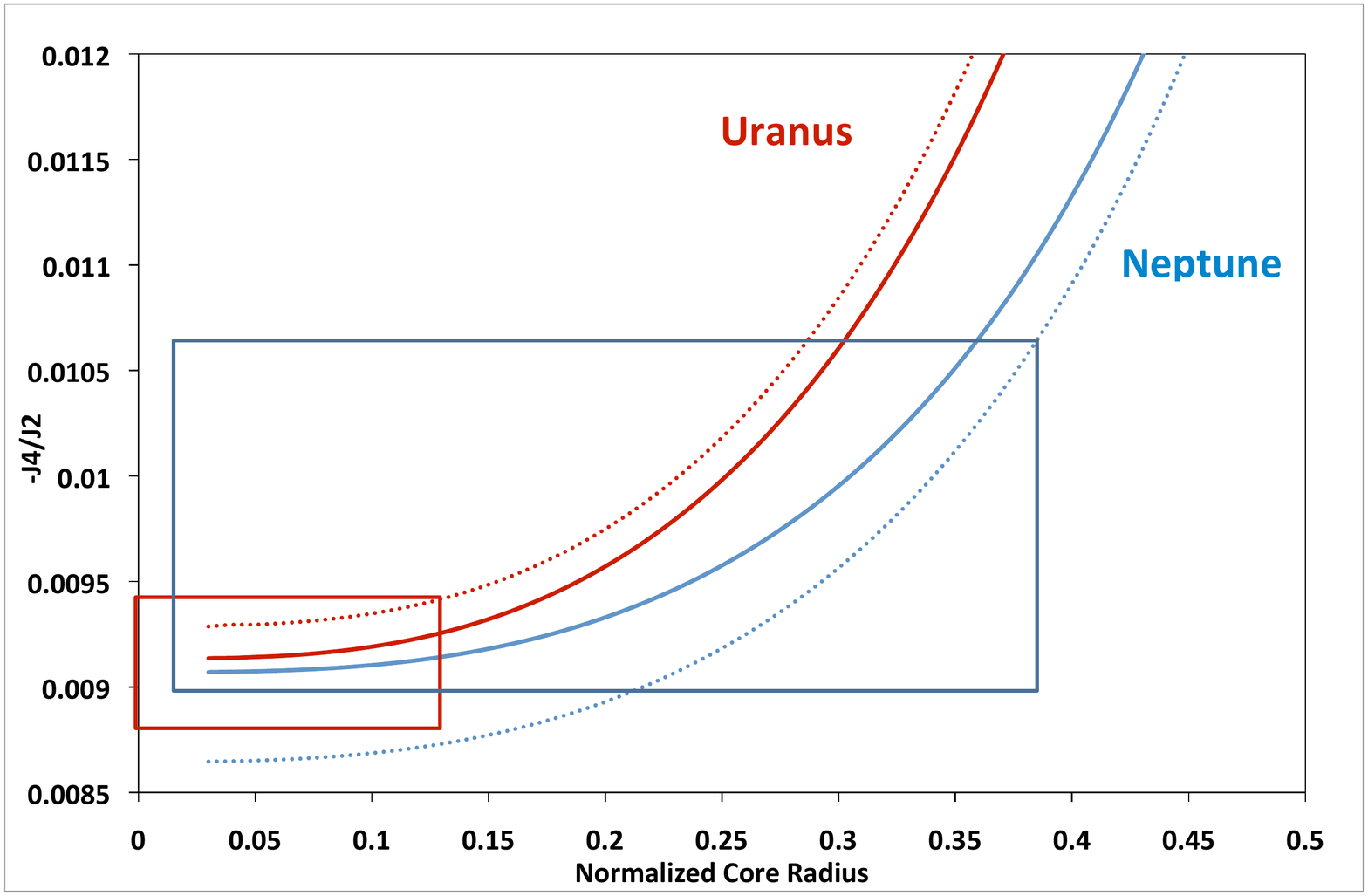}}
\caption{$-J_4/J_2$ vs. normalized core radius for Uranus (red) and Neptune (blue).  Solid curves are for Voyager rotation periods, dotted curves are for HAS10 periods.  The rectangles show the range of models with moments that match the observed values.}
\label{j2j4}
\end{figure}

For Uranus and Neptune the central mass concentration occupies a large enough volume to contribute directly to $J_2$ (Fig.\,(1) in Helled et al. 2011).  In some models it is large enough to contribute to $J_4$ as well.  As a result, the ratio $|J_4/J_2|$ contains information on the core's size.  We ran models of Uranus and Neptune assuming that they consist of a constant density core surrounded by an envelope where the density is given by a 6th order polynomial in radius (e.g. Schubert et al. 2012). The core density is set to 10\,g\,cm$^{-3}$, which is appropriate for rocky material under the expected conditions. The polynomial coefficients are chosen so that the radius, mass, and $J_2$ are reproduced to within the measured uncertainties.  $J_4$ is then computed for the model.  The results are shown in Fig.\,(\ref{j2j4}), where the ratio $\gamma\equiv-J_4/J_2$ is plotted vs. the core radius (normalized to the planet's radius). 

These models show that $\gamma$ increases with increasing core size. For the Voyager periods (solid) Uranus' core can be as large as $\sim 13\%$ of the planetary radius.  Neptune's core is less constrained because of the larger error bars on the measured moments, and may be substantially larger. For HAS10 periods (dotted) Uranus' core size is essentially unaffected, while Neptune's core must be even larger.  In fact, for these models, there is actually a lower limit to the size of Neptune's core. While detailed conclusions about the core size cannot be derived from these simple models, they do show how the interior structure is affected. Recent, more detailed models indeed show that the modified rotation periods give more centrally condensed models for Uranus than for Neptune (Nettelmann et al. 2012).

\section{Effect of Giant Impacts}

As mentioned earlier, a striking difference between Uranus and Neptune is Uranus' large axial tilt, and it is tempting to speculate that the giant impact which presumably caused the tilt (Safronov 1966) may also be responsible for the different internal structures and heat fluxes of the two planets (Stevenson 1986). To investigate the effect of the planet's tilt due to giant impacts we adapted a code that follows the motion of planetesimals through a protoplanetary envelope (Podolak et al. 1988). Although this code doesn't model the details of hydrodynamic effects or the high temperature/density equation of state effects, it can estimate the energy and angular momentum budget when giant impact occurs. For the planetary target we adapt one of the models of Podolak et al. (1995) with a rock core surrounded by an ice shell and gas envelope.  The details of the model aren't important, since we don't know what the pre-impact Uranus and Neptune looked like.  For our purposes the important thing is that the ice shell extends to 0.75 of the planetary radius which is typical of Uranus and Neptune models.  At this point there is a sharp density discontinuity between the outer envelope and the ice shell.  Large planetesimals that penetrate to this region will break up due to the sharp increase in ram pressure when they hit the shell.

\begin{figure}
\centerline{\includegraphics[width=14.2cm]{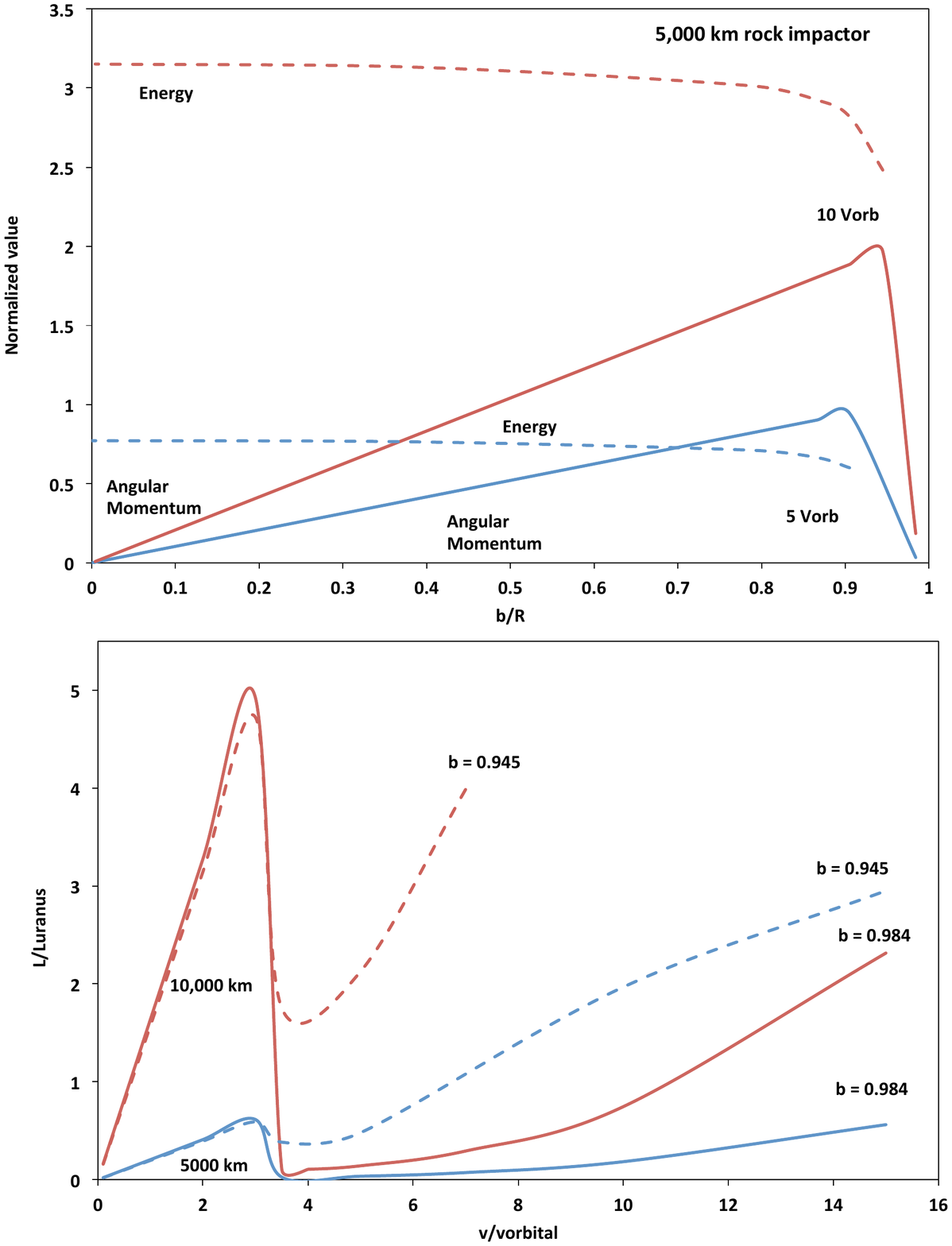}}
\caption{Upper panel: Angular momentum (solid) and energy (dashed) deposited vs.  impact parameter for $v=5v_{orbit}$ (blue) and $v=10v_{orbit}$ (red). Lower Panel: Angular momentum deposited as a function of impact velocity for a $10^4$\,km radius rock impactor (red) and a $5\times 10^3$\,km impactor (blue) for impact parameters of 0.984 $R_{Uranus}$ (solid) and 0.945 $R_{Uranus}$ (dashed).}
\label{impact1}
\end{figure}

The lower panel of Fig.\,\ref{impact1} shows the angular momentum deposition by a $5\times 10^3$\,km rocky body (blue) for impact parameters of $b=0.984\,R_{Uranus}$ (solid) and $b=0.945\,R_{Uranus}$ (dashed), respectively.  The angular momentum deposition is given in units of Uranus' present angular momentum, while the velocity is given in units of Uranus' orbital velocity. The escape velocity from Uranus is roughly three times its orbital velocity, so typical encounter velocities should be around $3v_{orbit}\lesssim v\lesssim 6v_{orbit}$ although special circumstances could lead to higher/lower velocities. 

As suggested from the figure there are two distinct regimes. For velocities $\lesssim 3.5v_{orbit}$ the planetesimal has a trajectory that allows it to penetrate into deep enough regions where the ram pressure forces exceed the internal strength of the material as well as its gravitational binding, and the planetesimal breaks up in the envelope.  In that case the planetesimal deposits all of its angular momentum in the planet. For higher velocities the planetesimal's trajectory keeps it in the upper envelope where the density is significantly lower and the planetesimal simply passes through the planet, causing a sharp drop in the angular momentum deposition. As the velocity increases still further the planetesimal leaves more of its angular momentum behind as a result of drag forces. For the smaller impact parameter a $5\times 10^3$\,km planetesimal can deposit sufficient angular momentum to change Uranus' tilt without being captured. For a $10^4$\,km planetesimal (red) the behavior is very similar, but the angular momentum deposited in the planet is correspondingly higher, and even at an impact velocity of $4v_{orbit}$ the planetesimal can pass through the planetary envelope and still deposit enough angular momentum to account for the tilt.

The upper panel of Fig.\,\ref{impact1} presents the energy (dashed) and angular momentum (solid) deposited vs. impact parameter for a $5\times 10^3$\,km planetesimal impacting the planet. The energy is normalized to the gravitational energy difference between a fully mixed planet and a differentiated planet. The angular momentum is normalized to Uranus' current angular momentum, and the impact parameter is normalized to the planet's radius. The blue and red curves correspond to impact velocities of $5v_{orbit}$ and $10v_{orbit}$, respectively. It is found that the angular momentum deposited increases as the impact parameter increases until some maximum value of $b$.  After that the planetesimal passes through the planet leaving a relatively small fraction of its angular momentum behind. The energy deposited at the point of breakup increases as the impact parameter decreases, and even for the case of $v=5v_{orbit}$ is somewhat $\lesssim1$ in these normalized units so that there's sufficient energy to mix large fractions of the core into the envelope.

Uranus and Neptune could have both suffered collisions with large bodies. Suppose that Uranus' collision was oblique, while Neptune's collision was more radial.  In the case of Uranus, the oblique collision could not only cause a large tilt in Uranus' spin axis, but could also deposit its energy into a thin shell to inhibit convection (Stevenson 1986).  A more radial collision for Neptune would mix the interior, and make the temperature profile closer to adiabatic, leading to more rapid cooling.  This could explain why Neptune has a strong internal heat source and Uranus doesn't.  It also explains the difference in the  cooling histories of Uranus and Neptune which was recently found to be even more prominent when using updated model atmospheres (Fortney et al. 2011). Indeed, our simulations show that radial collisions, which add relatively little angular momentum to the planet can have sufficient energy to mix large fractions of the core, while oblique collisions can add large amounts of angular momentum and still allow the core to survive.  Thus Uranus' large axial tilt may be directly related to the fact that it is more centrally condensed than Neptune.  This could be tested in the future by more careful measurements of the gravitational moments of Uranus and Neptune.  \par

\section{Summary}
The three measured parameters that are used to determine the structure of Uranus and Neptune's ($J_2, q, \omega$) are not self-consistent and any chosen pair predicts a different inertia factor for the planets.  The choice of $q$ and $J_2$ which are the ones used in interior modeling, gives very similar inertia factors for Uranus and Neptune implying that their internal structures are alike. However, when we use the HAS10  rotation periods, the parameters $q$, $f$ and $J_2$ are more self-consistent and Uranus is found to be more centrally condensed than Neptune. As a result, Uranus and Neptune may differ substantially in their internal structures (Nettelmann et al., 2012).

Two important results derive from our study.  First, one must be cautious in drawing conclusions about planetary structure simply on the basis of a mass classification.  Uranus and Neptune differ by only 10\% in mass and only 3\% in radius, but they may have very different interiors.  Second, these differences may be caused by secondary effects, such as the impact parameter of a giant impact.  These effects have important consequences for the classification of intermediate-mass bodies, and more detailed studies are necessary in order to evaluate the differences we might expect in the structure and composition of the many intermediate-mass exoplanets.

\section*{Acknowledgments}
We thank Jonathan Fortney for valuable comments. This work was supported, in part, by ISF grant 1231/10. 

\section*{References}

Anderson, J.~D. \& Schubert, G. 2007, Science, 317, 1384\\
{Bou\'{e}}, G. \& {Laskar}, J. 2010. ApJ,  712, L44\\
{Fortney}, J.~J. \& {Nettelmann}, N. 2010, Sp. Sci. Rev., 152, 423\\
{Fortney}, J. J., {Ikoma}, M., {Nettelmann}, N., {Guillot}, T., {Marley}, M. S. 2011. ApJ, 729, 32\\
{French}, R.~G., {Jones}, T.~J., \& {Hyland}, A.~R. 1987, Icarus, 69, 499\\
{Guillot}, T. 2005, Ann. Rev. Earth and Planet. Sci., 33, 493\\
Gurnett, D.~A., Lecacheux, A., Kurth, W.~S., Persoon, A.~M., Groene, J.~B.,
  Lamy, L., Z.~P., \& Carbary, J.~F. 2010, GRL, 36, 16102\\
Gurnett, D.~A., Persoon, A.~M., Kurth, W.~S., Groene, J.~B., Averkamp, T.~F.,
  Dougherty, M.~K., \& Southwood, D.~J. 2007, Science, 316, 442\\
Helled, R., Anderson, J.~D., Podolak, M., \& Schubert, G. 2011, ApJ,
  726, 1\\
{Helled}, R., {Anderson}, J.~D., \& {Schubert}, G. 2010, HAS10, Icarus, 210, 446\\
{Jacobson}, R.~A. 2007, B.A.A.S., 38, 4536\\
---. 2009, Astron. J., 137, 4322\\
{Karkoschka}, E. 2011. Icarus, 215, 439 \\
Lindal, G.~F. 1992, Astron. J., 103, 967\\
Lindal, G.~F., Lyons, J.~R., Sweetnam, D.~N., Eshleman, V.~R., \& Hinson, D.~P.
  1987, J. Geophys. Res., 14, 14,987\\
Marley, M., Gomez, P., \& Podolak, M. 1995, J. Geophys. Res., 100, 23349\\
{Ness}, N.~F., {Acuna}, M.~H., {Behannon}, K.~W., {Burlaga}, L.~F.,
  {Connerney}, J.~E.~P., \& {Lepping}, R.~P. 1986, Science, 233, 85\\
{Ness}, N.~F., {Acuna}, M.~H., {Burlaga}, L.~F., {Connerney}, J.~E.~P., \&
  {Lepping}, R.~P. 1989, Science, 246, 1473\\
Nettelmann, N., Helled, R., Fortney, J.~J., \& Redmer, R. 2012, Planet. and Sp.
  Sci., in press, 	arXiv:1207.2309v1\\
{Pearl}, J.~C. \& {Conrath}, B.~J. 1991, JGR, 96, 18921\\
Pearl, J.~C., Conrath, B.~J., Hanel, R.~A., Pirraglia, J.~A., \& Coustenis, A.
  1990, Icarus, 84, 12\\
Podolak, M., Pollack, J.~B., \& Reynolds, R.~T. 1988, Icarus, 73, 163\\
Podolak, M., Weizman, A., \& Marley, M. 1995, Planet. \& Space Sci., 43, 1517\\
Radau, R. 1885, Comptes Rendus, 100, 972\\
Redmer, R., Mattsson, T.~R., Nettelmann, N., \& French, M. 2011, Icarus, 211,
  798\\
{Safronov}, V.~S. 1966, Sov. Astron., 9, 987\\
{Stanley}, S. \& {Bloxham}, J. 2006, Icarus, 184, 556\\
{Stevenson}, D.~J. 1986, Lunar and Planetary Inst. Technical Report, 17, 1011\\
{Yoder}, C.~F. 1995, in Global Earth Physics: A Handbook of Physical Constants,
  ed. T.~J. {Ahrens}\\
Zharkov, V.~N. \& Trubitsyn, V.~P. 1978, Physics of {P}lanetary {I}nteriors, (Tucson: Pachart Press).\\

\begin{table}[h]
\caption{{\bf Input Parameters}}
\begin{tabular}{|c|c|c|c|c|}
\hline
Parameter & Jupiter & Saturn & Uranus & Neptune\\
\hline
Reference Radius & 71492 & 60330 & 26200 & 25225\\
Period & 9.925\,h & 10.570\,h & 17.240\,h & 16.11\,h\\
$f$ & $0.06487$ & $0.09796$ & $0.0229$ & $0.0171$ \\
$J_2\times 10^6$ & $14696.43\pm 0.21$ & $16290.71\pm 0.27$ & $3341.29\pm 0.72$ & $3408.43\pm 4.5$ \\
\hline
\end{tabular}
\end{table}

\begin{table}[h]
\caption{{\bf Inertia Factor} (The letters {\it V} and {\it H} in parenthesis refer to the Voyager and HAS10 periods, respectively.)}
\begin{tabular}{|c|c|c|c|c|c|c|}
\hline
Jupiter & Saturn & Uranus (V) & Neptune (V) & Uranus (H) & Neptune (H) & Method\\
\hline
0.2661 & 0.2278 & 0.2527 & 0.2067 & 0.2293 & 0.2549 & $q+f$\\
0.2658 & 0.2260 & 0.2219 & 0.2342 & 0.2161 & 0.2482 & $q+J_2$\\
0.2656 & 0.2254 & 0.2122 & 0.2452 & 0.2122 & 0.2452 & $f+J_2$\\
\hline
\end{tabular}
\end{table}

\end{document}